\documentclass{article}

\usepackage[margin=1in]{geometry}
\usepackage{graphicx}
\usepackage{amsmath}
\usepackage{amssymb}
\usepackage{hyperref}
\usepackage{booktabs}
\usepackage{multirow}
\usepackage{tabularx}
\usepackage{url}
\usepackage[utf8]{inputenc}
\usepackage[T1]{fontenc}
\usepackage{natbib}
\usepackage{authblk}
\usepackage{abstract}

\title{\textbf{XR-CareerAssist: An Immersive Platform for Personalised Career Guidance Leveraging Extended Reality and Multimodal AI}}

\author[1]{N.~D.~Tantaroudas\thanks{Corresponding author: nikolaos.tantaroudas@iccs.gr}}
\author[2]{A.~J.~McCracken}
\author[3]{I.~Karachalios}
\author[4]{E.~Papatheou}
\author[5]{V.~Pastrikakis}

\affil[1]{Institute of Communications and Computer Systems (ICCS), Iroon Polytechneiou 9, 15773, Zografou, Athens, Greece}
\affil[2]{DASKALOS-APPS, 183 Rue de l'Abb\'{e} Griffon, 01960 P\'{e}ronnas, France}
\affil[3]{National Technical University of Athens, Leof. Alimou, Katechaki, Zografou, 15772, Athens, Greece}
\affil[4]{Exeter Small-Scale Robotics Laboratory, Engineering Department, University of Exeter, Exeter, EX4 4QF, UK}
\affil[5]{CVCOSMOS Ltd, 1 Barnfield Crescent, Exeter, England, EX1 1QT}

\date{}

\begin{document}

\maketitle

\begin{abstract}
\noindent\textbf{Background:} Conventional career guidance platforms have long depended on static, text-driven interfaces that lack the capacity to engage users or deliver personalised, evidence-based insights. The pace of technological transformation calls for more advanced methodologies capable of assisting individuals as they traverse progressively complex professional landscapes. Computer-Assisted Career Guidance Systems (CACGS), although having undergone significant development since their origins in the 1960s, continue to be hampered by restricted interactivity and limited attention to the narrative aspects of career development.

\noindent\textbf{Methods:} This work introduces XR-CareerAssist, a novel platform that unifies Extended Reality (XR) with several Artificial Intelligence (AI) modules to provide immersive, multilingual career guidance. The system incorporates Automatic Speech Recognition (ASR) for voice-driven interaction, Neural Machine Translation (NMT) supporting English, Greek, French, and Italian, a Conversational Agent (Training Assistant) for personalised dialogue built on the Langchain framework, Vision-Language (VL) models fine-tuned on career visualisations employing BLIP with k-means clustering on representative Sankey diagram images, and Text-to-Speech (TTS) through AWS Polly for natural spoken responses delivered via an interactive 3D avatar. Career trajectories are depicted through dynamic Sankey diagrams derived from a repository of more than 100,000 anonymised professional profiles. The platform was built using Unity 2022.3 LTS with Meta SDK 2.0 targeting Meta Quest 3, with backend services hosted on AWS Elastic Beanstalk. A pilot evaluation was carried out at the University of Exeter involving 23 participants through structured questionnaires and semi-structured interviews.

\noindent\textbf{Results:} The pilot evaluation yielded robust validation outcomes including 95.6\% speech recognition accuracy, 78.3\% overall user satisfaction, and 91.3\% favourable ratings for system responsiveness. The platform architecture utilises AWS cloud infrastructure with FastAPI microservices, attaining response times below 200 milliseconds for career map generation. Stress testing confirmed scalability to 10,000 simultaneous users at 900\textendash 1,000 requests per second with zero failures. User feedback informed the complete implementation of enhancements addressing motion comfort, audio clarity, physical boundaries, and text legibility.

\noindent\textbf{Conclusions:} XR-CareerAssist constitutes a notable advance in career guidance technology, illustrating how the fusion of XR and AI can yield more engaging, accessible, and effective career development instruments. The successful unification of five AI modules within an immersive XR setting produces a multimodal interaction experience that sets XR-CareerAssist apart from current career guidance platforms. An earlier version of this work was presented at the XR Salento 2025 International Conference on Extended Reality~\citep{tantaroudas2026xrsalento}. The platform builds upon related work on AI-powered XR systems for inclusive communication~\citep{tantaroudas2026interact} and AI-based services for language learning in immersive environments~\citep{tantaroudas2026ailanguage}.

\end{abstract}

\section{Introduction}
\label{sec:introduction}

Professional development in today's labour market poses unprecedented difficulties as individuals must navigate an increasingly intricate landscape defined by swift technological evolution, globalisation, and shifting skill demands~\citep{hirschi2018}. Conventional career guidance methods, grounded in trait-factor matching theories originating from Parsons' foundational work in 1909, have progressed substantially yet still encounter core limitations regarding engagement, personalisation, and accessibility~\citep{maree2013}. Computer-Assisted Career Guidance Systems (CACGS) appeared in the 1960s as technological means to broaden the reach of career counselling, but numerous current implementations continue to rely on text-based interfaces, offer restricted interactivity, and pay inadequate attention to the narrative and contextual aspects of career development~\citep{leung2022}.

The theoretical underpinning for contemporary career guidance has moved considerably toward constructivist perspectives, notably Career Construction Theory (CCT), which stresses the subjective and narrative character of career development~\citep{savickas2012}. CCT treats individuals as active narrators of their career stories rather than passive receivers of advice, implying that effective career interventions ought to promote self-reflection, meaning-making, and the formation of personally significant career narratives. This perspective carries important ramifications for the design of career guidance technologies, necessitating systems that support exploration, visualisation, and dialogue rather than merely delivering information or matching profiles to particular roles.

Extended Reality (XR) technologies, encompassing Virtual Reality (VR), Augmented Reality (AR), and Mixed Reality (MR), present transformative possibilities for career guidance by establishing immersive settings that foster experiential learning and exploration~\citep{garcia2021}. Studies have shown that XR can heighten engagement, strengthen information retention, and forge emotional connections unattainable through text-based systems~\citep{shepiliev2021}. Nevertheless, the deployment of XR for career guidance remains at an early stage, with sparse empirical investigation into its efficacy and even fewer implementations that fully exploit the multimodal interaction capabilities these technologies provide.

AI components such as Automatic Speech Recognition (ASR), Neural Machine Translation (NMT), and conversational agents have reached significant maturity, enabling natural language interactions that can approximate human dialogue~\citep{bommasani2021}. When coupled with XR, these technologies open avenues for career guidance systems that respond in the user's preferred language, sustain contextual conversations, and offer personalised insights informed by individual profiles and preferences. Recent advances in AI-driven XR platforms for inclusive communication~\citep{tantaroudas2026interact} and immersive language learning~\citep{tantaroudas2026ailanguage} further demonstrate the potential of combining these modalities for user-centred applications.

An initial version of this work was presented at the XR Salento 2025 International Conference on Extended Reality~\citep{tantaroudas2026xrsalento}, where the conceptual architecture and preliminary design specifications for XR-CareerAssist were introduced. The present paper significantly extends that work by reporting the full technical implementation across five separate AI modules (ASR, NMT, VL, Conversational Agent, and TTS), comprehensive pilot evaluation results from 23 participants at the University of Exeter, quantitative performance metrics, qualitative user feedback analysis, and an in-depth discussion of ramifications for the domain of AI-enhanced career guidance systems.

The system tackles the shortcomings of existing approaches through several principal innovations. First, XR-CareerAssist establishes an immersive XR setting that allows users to investigate career paths via interactive visualisations and spatial interfaces instead of static web pages. Second, the platform delivers multimodal AI integration through the cohesive combination of five AI technologies: ASR for voice input, NMT for multilingual support, a Conversational Agent (Training Assistant) for contextual dialogue, Vision-Language models for visual interpretation, and TTS for natural spoken responses. Third, personalised career mapping using dynamic Sankey diagrams produced from over 100,000 anonymised professional profiles lets users visualise prospective career trajectories grounded in their personal attributes. Fourth, interaction through a 3D avatar interface offers an interactive virtual assistant that engages users in natural conversation, reinforcing the sense of presence and personal connection. The platform was built following an agile methodology, advancing from concept (TRL3, TRL4) to a validated prototype demonstrated in a realistic setting (TRL6).

The remainder of this paper is organised as follows: Section~\ref{sec:related} surveys related work in CACGS, XR applications in education and career guidance, and multimodal AI systems. Section~\ref{sec:architecture} describes the system architecture, encompassing AI component integration and the technical implementation. Section~\ref{sec:pilot} presents the pilot study methodology and comprehensive results. Section~\ref{sec:discussion} discusses findings, implications, and limitations. Section~\ref{sec:conclusions} concludes with future research directions.

\section{Related Work and Background}
\label{sec:related}

\subsection{Development of Computer-Assisted Career Guidance Systems}

The origins of CACGS date back to the 1960s when the first systems sought to employ computers in career counselling~\citep{sampson2000}. These pioneering systems primarily concentrated on aligning individual attributes with occupational demands, mirroring the prevailing trait-factor paradigm of that period. This approach presumed that optimal career choices arose from matching personal traits (interests, capabilities, values) with job specifications, a viewpoint that, while methodical, frequently neglected the dynamic and constructive character of career development~\citep{brown2016}.

Present-day CACGS have progressed beyond straightforward matching to encompass more advanced features including career exploration instruments, labour market information databases, and decision-support frameworks~\citep{leung2022}. Prominent examples include O*NET (Occupational Information Network), which supplies thorough occupational data, and career exploration platforms provided by leading job search engines. However, recent investigations identify enduring shortcomings in current CACGS, including inadequate incorporation of Career Construction Theory principles, constrained personalisation beyond surface-level matching, and interfaces that do not engage users in substantive self-reflection and narrative construction~\citep{leung2022}.

The C3-IoC system created by Jos\'{e}-Garc\'{i}a et al.~\citep{josgarcia2023} exemplifies recent progress in applying machine learning to career guidance. Their method employs network visualisation and machine learning algorithms to evaluate student competencies and propose career pathways, showing how AI can augment the analytical capabilities of career systems. Nevertheless, such systems generally remain limited to conventional screen-based interfaces, missing the chance to harness immersive technologies for heightened engagement and experiential learning.

\subsection{Extended Reality in Education and Career Guidance}

XR technologies have exhibited considerable potential in educational settings, with research consistently demonstrating advantages for engagement, motivation, and learning outcomes, especially for complex or abstract concepts amenable to visualisation~\citep{radianti2020}. The immersive quality of XR facilitates experiential learning approaches that would be impractical or unachievable in traditional environments, spanning from virtual laboratory experiments to historical reconstructions.

Uses of XR specifically for career guidance remain comparatively scarce but yield encouraging outcomes. Garcia Estrada and Prasolova-F{\o}rland~\citep{garcia2021} created VR content for career guidance during the COVID-19 pandemic, discovering that immersive experiences could partly offset the absence of face-to-face guidance opportunities. Their work established that VR could successfully communicate occupational information and support career exploration, although they highlighted the necessity for more advanced AI integration to enable genuinely personalised experiences.

Shepiliev et al.~\citep{shepiliev2021} investigated WebAR (Augmented Reality accessible via web browsers) for career guidance quests, developing gamified experiences that steer users through career exploration tasks. Their approach demonstrated greater engagement relative to conventional methods, though the AR implementation was inherently restricted by browser capabilities and lacked the natural language processing or conversational AI that would permit more dynamic interactions.

Earlier research on immersive learning more generally indicates that VR environments can improve spatial comprehension, boost retention of procedural knowledge, and establish emotional connections that affect decision-making~\citep{jensen2018}. For career guidance, these capabilities suggest potential advantages in helping users visualise career paths, grasp workplace environments, and cultivate emotional investment in career objectives. However, achieving these advantages demands careful focus on user experience design, technical performance, and alignment with pedagogically sound guidance methodologies.

\subsection{Multimodal AI for Natural Interaction}

The convergence of multiple AI technologies enables natural, multimodal interaction that more closely mirrors human communication. ASR systems have attained near-human precision for numerous languages and acoustic conditions~\citep{radford2023}, while NMT has reached quality levels appropriate for real-time cross-linguistic communication~\citep{fan2021}. Conversational AI systems, driven by large language models, can sustain contextual dialogue and deliver substantive responses across varied domains~\citep{brown2020}.

Voice-driven interaction provides particular benefits for XR environments, where conventional input devices (keyboards, mice) are unavailable or cumbersome. ASR permits hands-free interaction that preserves immersion while enabling users to articulate complex queries and participate in natural dialogue. When paired with NMT, systems can accommodate users irrespective of their native language, addressing a vital accessibility requirement for worldwide deployment~\citep{pratap2020}.

Vision-Language (VL) models constitute a more recent development, allowing systems to comprehend and characterise visual content~\citep{du2022}. These models can parse images, diagrams, and visualisations, potentially enabling career guidance systems to elucidate complex career trajectories rendered in visual formats. The BLIP (Bootstrapping Language-Image Pre-training) model~\citep{li2022blip} illustrates this capability, though domain-specific fine-tuning is generally necessary for specialised applications such as career diagram interpretation.

Text-to-Speech (TTS) technology closes the interaction loop, enabling systems to furnish spoken responses that improve naturalness and accessibility. Contemporary neural TTS systems, including cloud services such as AWS Polly, generate speech approaching human quality, although voice selection and prosody remain key design considerations~\citep{tan2021}.

\subsection{The CVCOSMOS Platform and VOXReality Ecosystem}

XR-CareerAssist is built upon two foundational platforms that supply essential capabilities for the career guidance system. The CVCOSMOS platform provides an extensive database of over 100,000 anonymised professional CVs, facilitating data-driven analysis of career trajectories across industries, roles, and timeframes~\citep{cvcosmos2024}. This database furnishes the empirical basis for producing personalised career maps, enabling users to observe how professionals with comparable backgrounds have advanced through their careers. The platform makes available APIs for querying career data and generating statistical analyses, rendering it well-suited for integration with front-end applications.

The VOXReality project is a Horizon Europe initiative centred on developing and validating immersive media technologies~\citep{voxreality2024}. VOXReality supplies the core AI models that drive the multimodal interaction capabilities of XR-CareerAssist, including ASR for speech recognition, NMT for translation, and LLMs for conversational AI. By extending these general-purpose models for career guidance applications, XR-CareerAssist capitalises on established technologies while tailoring them to domain-specific needs. The combination of CVCOSMOS career data with VOXReality AI models, delivered through an immersive XR interface, produces a distinctive integration that differentiates XR-CareerAssist from existing career guidance solutions.

\section{System Design and Architecture}
\label{sec:architecture}

\subsection{Architectural Overview}

\begin{figure}[htbp]
\centering
\includegraphics[width=\textwidth]{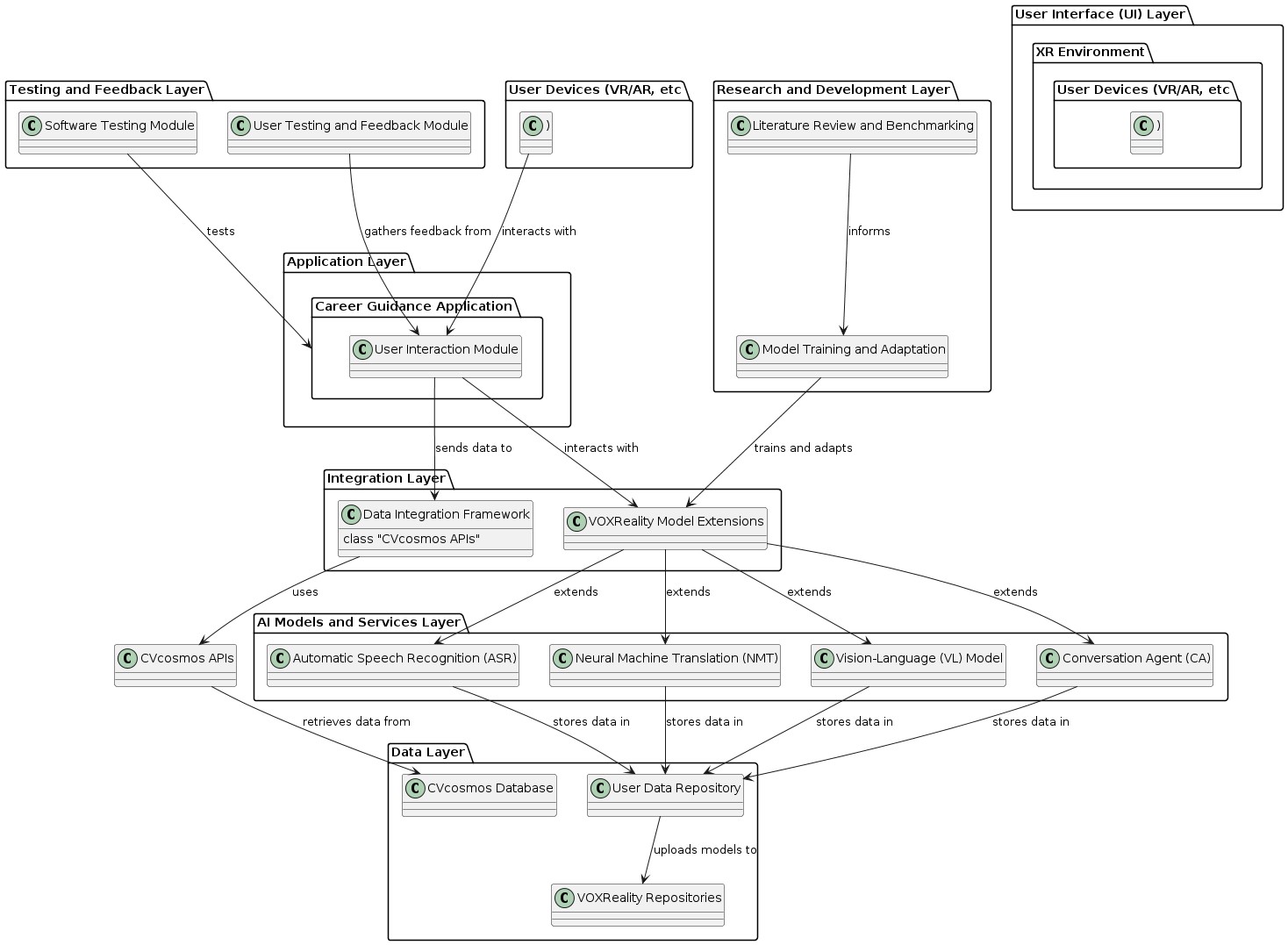}
\caption{XR-CareerAssist architecture diagram illustrating the layered system design: User Interface (UI) Layer with VR/AR devices; Application Layer with the Career Guidance Application; Integration Layer incorporating the Data Integration Framework (APIs) and VOXReality Model Extensions; AI Models and Services Layer hosting ASR, NMT, VL Model, and Conversational Agent (CA); and Data Layer comprising the CVs Database and User Data Repository linked to VOXReality Repositories.}
\label{fig:architecture}
\end{figure}

XR-CareerAssist utilises a layered architecture designed for modularity, scalability, and ease of maintenance. The system is composed of five principal layers: User Interface Layer, Application Layer, Integration Layer, AI Models and Services Layer, and Data Layer. Figure~\ref{fig:architecture} presents the complete system architecture depicting the relationships among these layers and the flow of data through the various AI components.

The User Interface Layer encompasses the immersive XR environment accessed via VR headsets, specifically the Meta Quest 3 platform. This layer manages user input through voice commands, gesture controls, and gaze-based selection, while rendering the 3D environment, career visualisations, and avatar interactions.

The Application Layer governs the core career guidance logic, including questionnaire administration, user session management, and coordination of AI model inference. This layer converts user inputs into suitable API calls and orchestrates the response pipeline from multiple AI components.

The Integration Layer offers abstraction between the application and external services, implementing adapters for the CVs database and AI models. This layer handles authentication, rate limiting, error management, and data transformation, ensuring dependable communication with backend services.

The AI Models and Services Layer houses the core AI capabilities: ASR for speech-to-text conversion, NMT for translation, the Conversational Agent for dialogue management, VL models for visual interpretation, and TTS for speech synthesis. Each component is independently deployable and scalable.

The Data Layer oversees persistent storage including user profiles, session data, and cached career information. The layer interfaces with the CVs database for career trajectory data while safeguarding user privacy through proper data handling protocols.

\subsection{Envisioned User Scenario}

\begin{figure}[htbp]
\centering
\includegraphics[width=\textwidth]{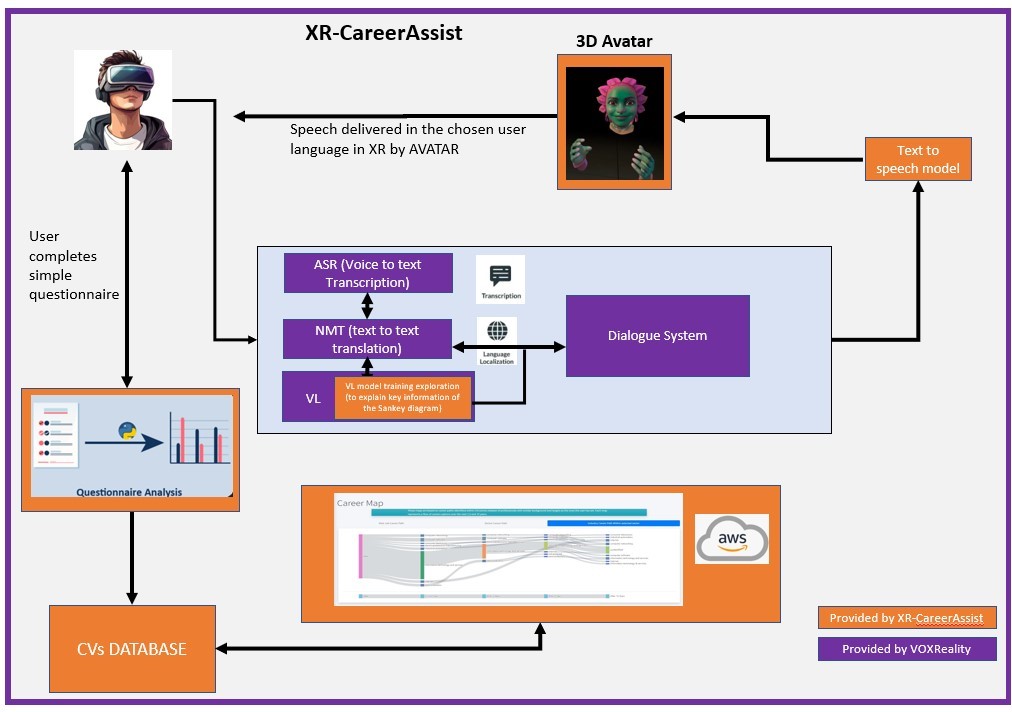}
\caption{Envisioned scenario of XR-CareerAssist showing the complete user journey and AI integration flow, including voice input processing through ASR, multilingual translation via NMT, career data retrieval, VL model interpretation, dialogue generation, and TTS audio output delivered through the 3D Avatar.}
\label{fig:scenario}
\end{figure}

Figure~\ref{fig:scenario} presents the complete envisioned scenario for XR-CareerAssist, illustrating how users engage with the system and how the different AI components collaborate to provide personalised career guidance. The user wearing a VR headset interacts with a 3D Avatar in the XR environment. The spoken query is processed through the ASR module for transcription. The NMT module translates the text if required for multilingual support. The system queries the CVs Database to obtain relevant career profiles. The VL model interprets career visualisations (Sankey diagrams) to supply explanations. The Dialogue System generates a personalised response, which is sent through the TTS service operating on AWS infrastructure to produce natural speech output delivered via the 3D Avatar.

Table~\ref{tab:journey} summarises the full user journey and AI model integration, indicating each interaction step and the corresponding AI component involved.

\begin{table}[htbp]
\centering
\caption{User journey and AI model integration within XR-CareerAssist.}
\label{tab:journey}
\footnotesize
\begin{tabular}{p{2.2cm} p{3.5cm} p{3.5cm} p{3.5cm}}
\toprule
\textbf{Step} & \textbf{User Input} & \textbf{Model/Process} & \textbf{Output} \\
\midrule
Questionnaire completion & Job role, experience, skills, target role, industry preferences & CVs database queried via API & Career profiles matched, Sankey diagrams generated \\
\addlinespace
Voice-based navigation & Voice queries and navigation commands & ASR Model: spoken input converted to text & Transcribed text for NMT processing \\
\addlinespace
Multilingual interaction & Text output from ASR & NMT Model: translates to user's chosen language & Multilingual accessibility enabled \\
\addlinespace
Career map exploration & User-requested insights for Sankey diagram & VL Model: interprets Sankey diagrams & Career progression explained in text \\
\addlinespace
Career questioning & User queries about career paths & Dialogue System: generates personalised advice & Tailored guidance delivered \\
\addlinespace
Listening to responses & Continued interaction & TTS: text converted to speech via 3D Avatar & Audio responses enhance engagement \\
\bottomrule
\end{tabular}
\end{table}

\subsection{Backend Infrastructure}

The backend services are deployed on AWS Elastic Beanstalk, running on a GPU-enabled EC2 instance (g4dn.xlarge with 16~GB GPU memory) on Ubuntu. The deployment comprises NGINX as a reverse proxy for efficient request handling and SSL termination, Python 3.12 runtime with FastAPI for high-performance API services, and DuckDB for in-memory analytical queries on career data.

The API architecture adheres to RESTful principles with endpoints for questionnaire submission (\texttt{/profile/sankey} POST), AI model invocations (\texttt{/asr}, \texttt{/nmt}, \texttt{/tts}, \texttt{/dialogue}), and health monitoring. Response caching minimises redundant processing, especially for commonly requested career scenarios.

\subsection{System Component Implementation and Validation}

\begin{figure}[htbp]
\centering
\includegraphics[width=\textwidth]{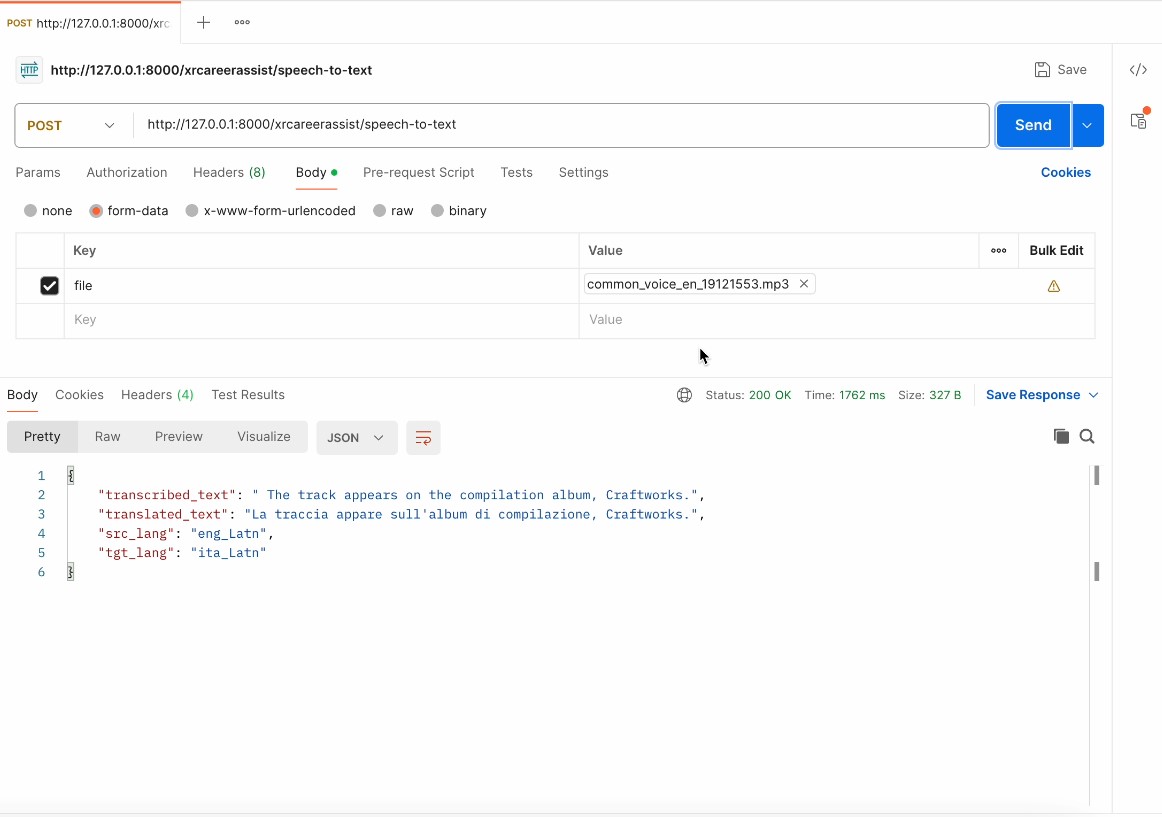}
\caption{XR ASR pipeline: English voice input to text transcription and NMT translation to Italian. The display includes the audio input waveform, ASR transcription output, and NMT translation with timing metrics, confidence scores, and language detection indicators.}
\label{fig:asr}
\end{figure}

\textbf{Automatic Speech Recognition (ASR):} The ASR component serves as the primary voice input channel for XR-CareerAssist. The system receives voice input directly from the XR environment, processes spoken queries through the ASR model supporting English, Greek, and French, and returns transcribed text with confidence scores. The model attains recognition accuracy surpassing 95\% in controlled settings, with response latency below 2 seconds for typical queries. Figure~\ref{fig:asr} depicts the ASR pipeline showing the complete flow from English voice input to text transcription, together with subsequent NMT translation to Italian.

\textbf{Neural Machine Translation (NMT):} The NMT component enables multilingual accessibility by translating text between supported language pairs. The system offers translation capabilities for English, Greek, French, and Italian, with bidirectional translation support and response latency below 2 seconds per translation request. Integration with ASR permits seamless translation of voice input, while coupling with TTS ensures translated responses can be spoken in the user's preferred language.

\begin{figure}[htbp]
\centering
\includegraphics[width=\textwidth]{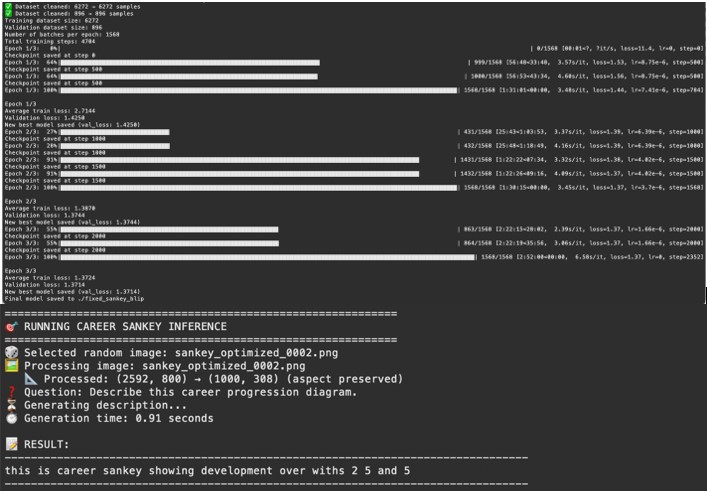}
\caption{VL model extended through fine-tuning to respond to Sankey diagram queries. Two examples demonstrate the fine-tuned BLIP model answering questions about career transition patterns and relative frequencies of different career moves.}
\label{fig:vlmodel}
\end{figure}

\textbf{Vision-Language Model (VL):} To enable natural language interpretation of career visualisations, the team fine-tuned a BLIP-based Vision-Language model. The fine-tuning procedure employed k-means clustering to select 20 representative Sankey diagram images from the career data corpus, followed by training for 50 epochs. The resultant model can address user queries regarding Sankey diagrams, explaining career transitions and patterns depicted in the visualisations. Figure~\ref{fig:vlmodel} illustrates the VL model's performance on sample Sankey diagram queries.

\begin{figure}[htbp]
\centering
\includegraphics[width=\textwidth]{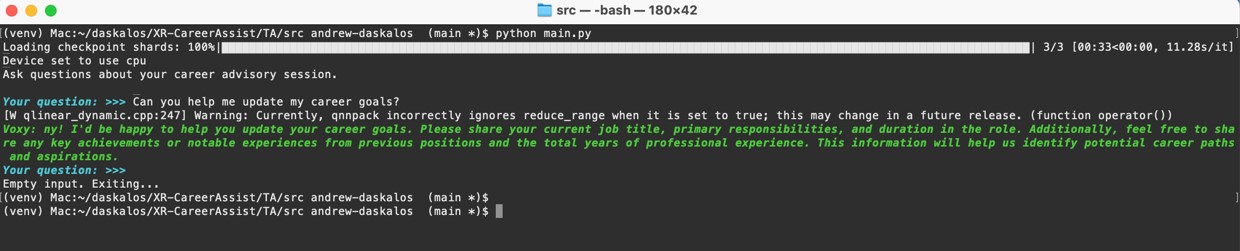}
\caption{Dialogue system training manual for XR-CareerAssist, providing structured guidance for interacting with the Training Assistant including step-by-step instructions, best practices for career-related questions, and troubleshooting information.}
\label{fig:dialogue}
\end{figure}

\textbf{Conversational Agent (Training Assistant):} The Conversational Agent provides contextual dialogue capabilities within the XR environment. Built on the Langchain framework, it maintains conversation context, interfaces with career data APIs for personalised responses, and escalates complex queries to external resources when necessary. It functions as the primary channel through which users engage with career guidance features, with its responses delivered through the 3D avatar via TTS. Figure~\ref{fig:dialogue} presents the dialogue system training manual that guides users in their interactions.

\begin{figure}[htbp]
\centering
\includegraphics[width=\textwidth]{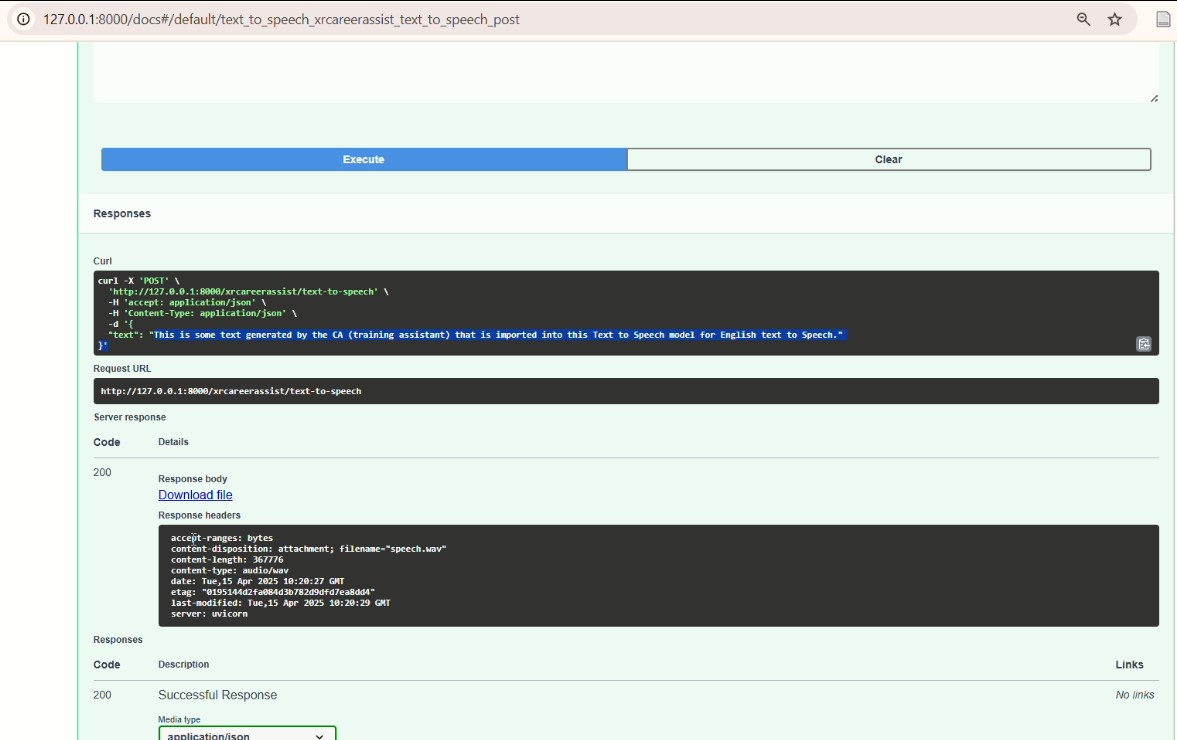}
\caption{Local testing of PIPER TTS with XR-CareerAssist backend, demonstrating the initial evaluation before transitioning to AWS Polly for production deployment.}
\label{fig:piper}
\end{figure}

\textbf{Text-to-Speech (TTS):} To enhance the immersive and accessible nature of XR-CareerAssist, a TTS feature was incorporated to convert generated textual outputs into realistic audio responses for users within the XR environment. Initially, the team evaluated the open-source PIPER TTS engine, which was successfully installed, configured, and tested locally as shown in Figure~\ref{fig:piper}. However, PIPER exhibited significant technical constraints including dependence on unmaintained Python libraries and incompatibility with Python 3.12 used throughout the XR-CareerAssist backend. As a substitute, the team deployed AWS Polly, a robust and secure cloud-based TTS service, providing high reliability, adherence to security best practices, and consistent performance. Multiple Polly voices were assessed and chosen to ensure natural-sounding speech aligned with the XR user experience objectives.

\begin{figure}[htbp]
\centering
\includegraphics[width=\textwidth]{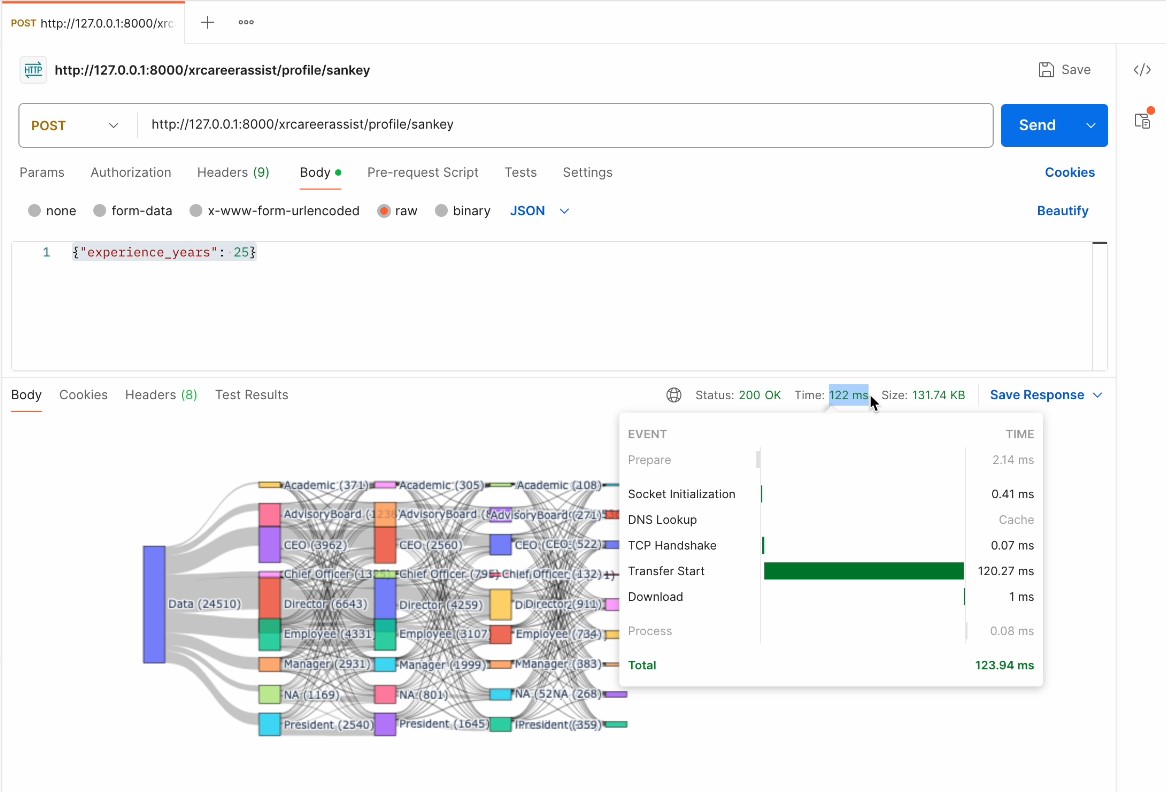}
\caption{Sample Sankey diagram generated from CVCOSMOS data for a user profile input with 25 years of experience. Flowing paths indicate career transitions from current roles through intermediate stages to potential future roles over a 10-year projection, with flow widths representing relative transition frequencies.}
\label{fig:sankey}
\end{figure}

\textbf{Career Data Integration and Visualisation:} A key technical achievement involved the successful integration of the CVs database with the XR-CareerAssist backend infrastructure. The project team employed DuckDB for efficient in-memory analytics, and a dedicated API service was implemented in Python to expose high-performance endpoints for querying the anonymised profiles dataset. The API endpoint (\texttt{/profile/sankey} POST) accepts questionnaire-style inputs and returns a PNG image of a generated Sankey diagram. Figure~\ref{fig:sankey} displays a sample Sankey diagram generated from CVCOSMOS data showing career transitions for a user profile with 25 years of experience spanning a 10-year projection period.

\begin{figure}[htbp]
\centering
\includegraphics[width=\textwidth]{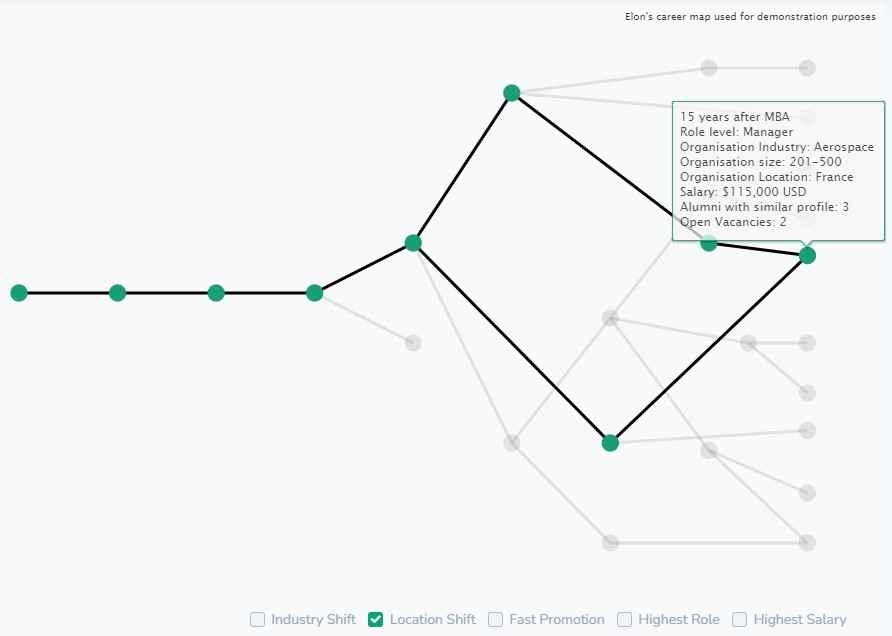}
\caption{Career map concept note for location shift target showing detailed metrics for career progression tracking.}
\label{fig:careermap}
\end{figure}

The career map concept illustrates how users can visualise their prospective career progression over time, following professional journeys including role advancements and industry transitions. Figure~\ref{fig:careermap} presents the Career Map concept note with detailed metrics.

\begin{figure}[htbp]
\centering
\includegraphics[width=\textwidth]{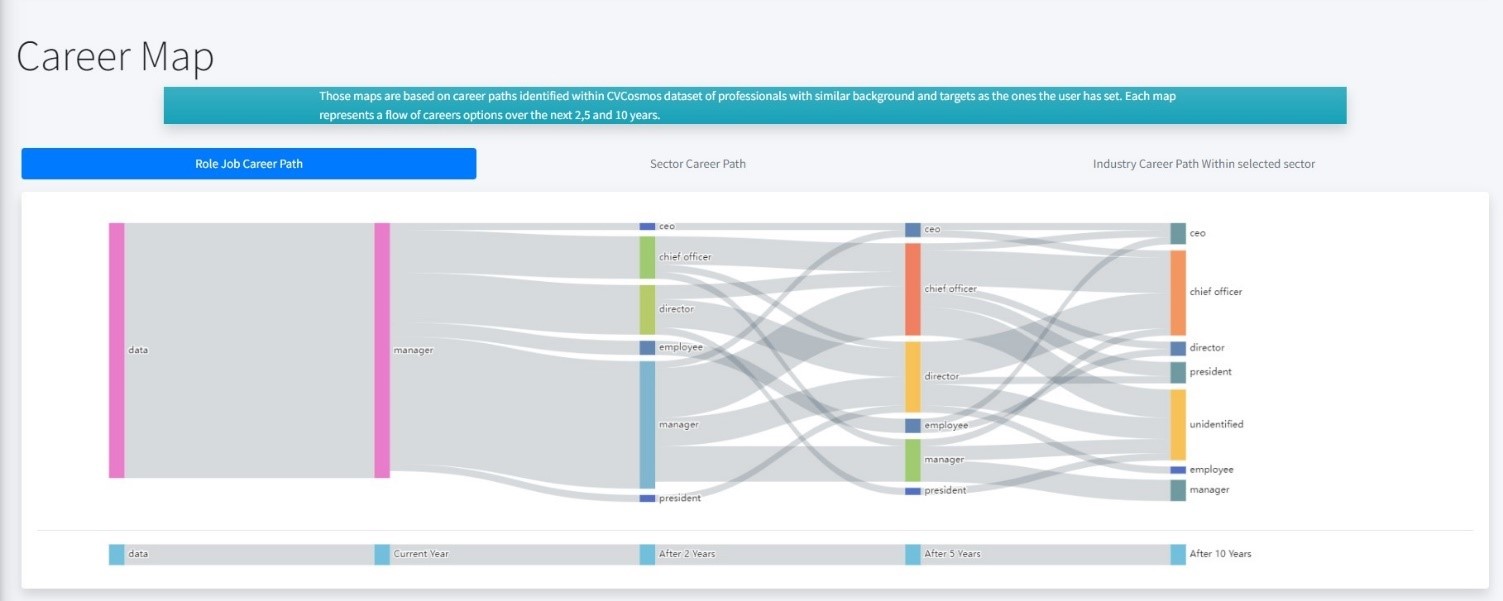}
\caption{Job role evolution career map for specified user input showing 10-year progression. The left column displays starting roles (Employee, Manager, Director), middle columns show intermediate career stages at 2 and 5 years, and the right column shows potential roles at 10 years with flow widths indicating transition frequencies.}
\label{fig:jobrole}
\end{figure}

\begin{figure}[htbp]
\centering
\includegraphics[width=\textwidth]{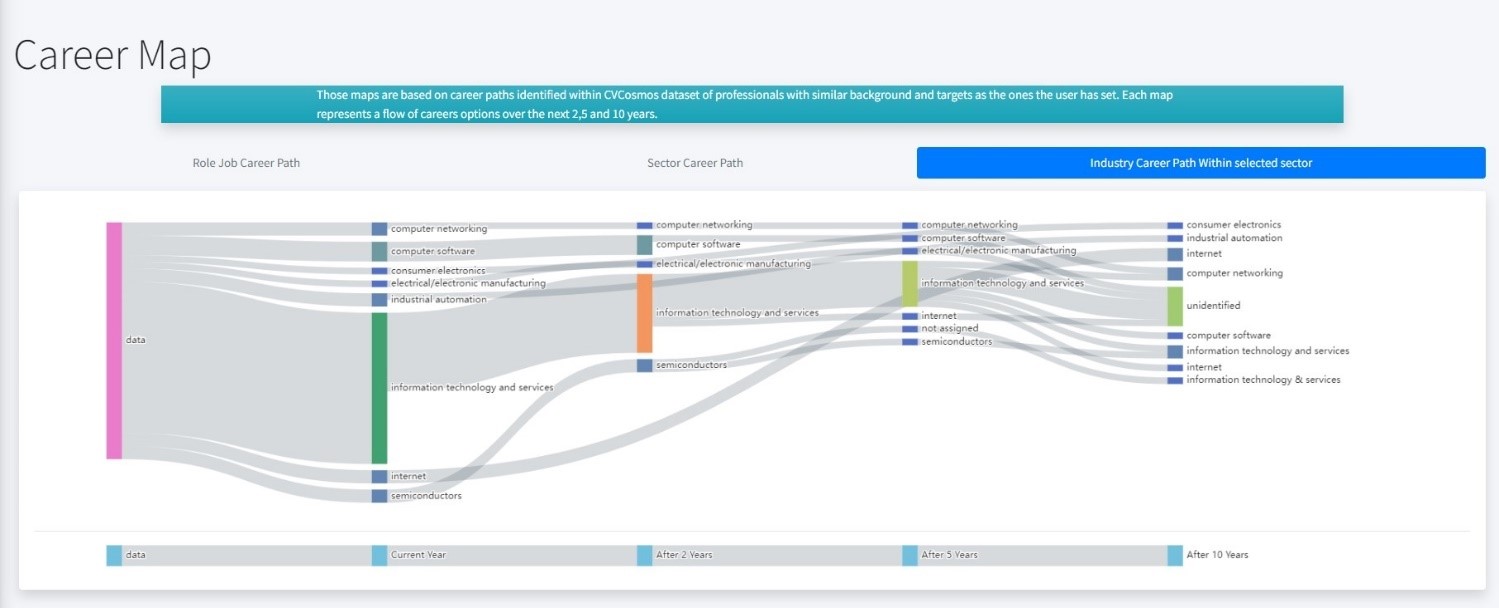}
\caption{Industry shift evolution career map for specified user input over 10 years, showing how professionals typically transition between industries with flow widths representing relative transition frequencies across sectors including Computer Manufacturing, Financial Services, IT Services, and Software Products.}
\label{fig:industryshift}
\end{figure}

The platform furnishes users with dynamic career maps displaying potential role and industry transitions across different time horizons (2, 5, and 10 years). Figure~\ref{fig:jobrole} shows the Job Role Evolution Career Map and Figure~\ref{fig:industryshift} presents the Industry Shift evolution diagram.

\textbf{XR User Interface and Experience:} The XR environment was developed with Unity 2022.3 LTS and Meta SDK 2.0, optimised for the Meta Quest 3 platform. The interface features interactive 3D elements that steer users through career exploration activities, with user-friendly navigation ensuring accessibility for individuals with differing levels of XR familiarity.

Career questionnaires were designed to gather user data on professional interests, competencies, and career aspirations. The design prioritises clarity and brevity to sustain user engagement, with questions tailored to elicit information essential for generating precise career transition visualisations.

\begin{figure}[htbp]
\centering
\includegraphics[width=\textwidth]{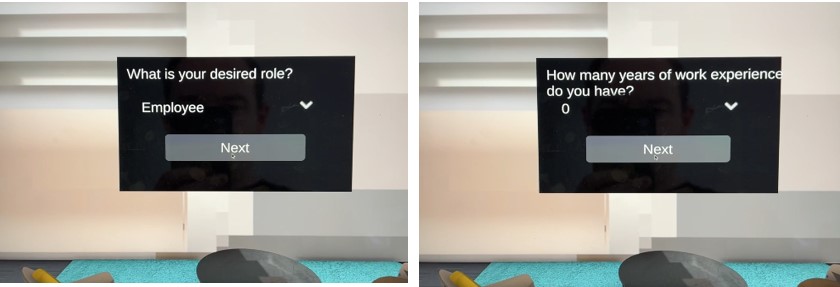}
\caption{Questionnaire flow within the XR environment showing two sequential input screens: role selection (left) and work experience specification (right), rendered in a clean, dark-themed design optimised for VR readability.}
\label{fig:questionnaire}
\end{figure}

Figure~\ref{fig:questionnaire} illustrates the questionnaire flow within the XR environment. Users specify their desired role via a dropdown menu offering career level options (Employee, Manager, Director, Executive) and indicate their years of work experience. Based on these inputs, the system queries the CVCOSMOS database containing over 100,000 anonymised CVs from experienced professionals to identify matching career trajectories and potential career shifts. The algorithm analyses profiles with comparable starting characteristics to produce personalised Sankey diagrams showing statistically derived career progression pathways across 2, 5, and 10-year time horizons.

\begin{figure}[htbp]
\centering
\includegraphics[width=\textwidth]{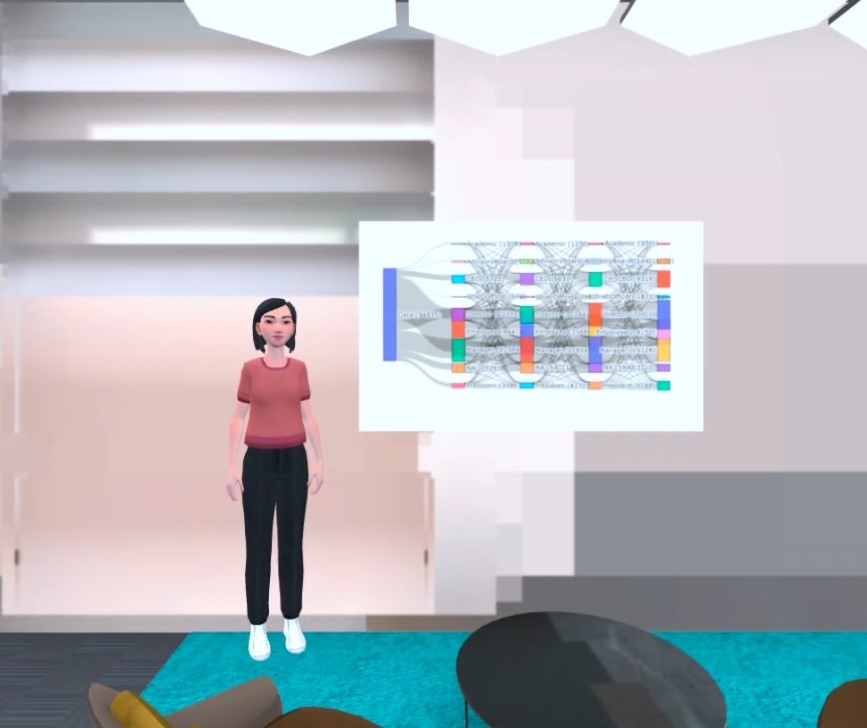}
\caption{XR-CareerAssist user interface within the XR environment as experienced on the Meta Quest 3 headset, showing the 3D avatar (Training Assistant) alongside the Sankey diagram display panel in the virtual workspace.}
\label{fig:xrinterface}
\end{figure}

Figure~\ref{fig:xrinterface} depicts the main XR interface where users interact with the career guidance features. The 3D avatar (Training Assistant) is positioned to the left, with the Sankey diagram panel prominently displayed for viewing and discussion within a modern, professional virtual environment.

Users can engage in natural voice-driven dialogue to inquire about the displayed career visualisations. When a user poses a question regarding the Sankey diagram, the integrated microphone captures the spoken query. The ASR module transcribes the audio input into text in real time. If the transcribed text is in a language other than English (e.g., Italian or French), the NMT module automatically translates it into English for downstream processing. The translated text is then forwarded to the Conversational Agent, which is specifically trained to interpret career-related queries and generate contextually appropriate textual responses. Finally, the generated response is converted into natural-sounding speech using Amazon Polly's TTS service, and the audio output is delivered through the 3D avatar, producing a seamless, immersive conversational experience that guides users through their personalised career exploration.

\textbf{Performance Optimisation:} The backend API, specifically the Sankey diagram generation endpoint, underwent extensive performance optimisation. Initial implementations required approximately 45 seconds to produce personalised career maps. Following optimisation through query restructuring, index tuning, caching strategies, and algorithm refinements, response times were reduced to roughly 200 milliseconds, constituting a 99.56\% improvement.

\begin{figure}[htbp]
\centering
\includegraphics[width=\textwidth]{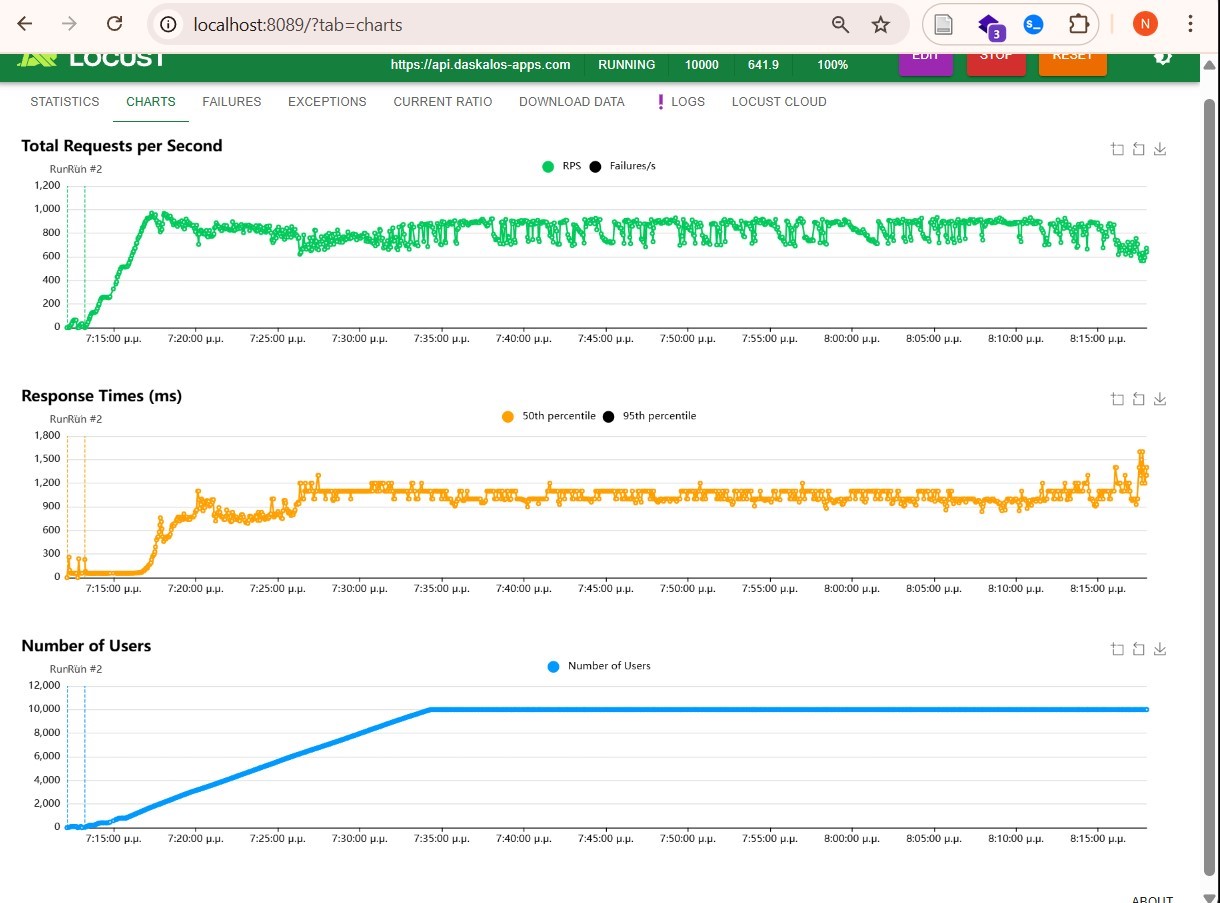}
\caption{Load testing results for 10,000 concurrent users. The top graph shows sustained throughput of 900 to 1,000 requests per second, the middle graph displays response time distribution with median latency between 800 and 900 milliseconds and 95th percentile under 2 seconds, and the bottom section shows zero failure rate throughout the test.}
\label{fig:loadtest}
\end{figure}

To confirm scalability requirements, thorough load testing was performed for 10,000 concurrent simulated users. Figure~\ref{fig:loadtest} displays the load testing results showing the platform's behaviour under heavy load.

\section{Pilot Study}
\label{sec:pilot}

\subsection{Methodology}

A pilot evaluation was conducted at the University of Exeter to assess XR-CareerAssist's functionality, usability, and user perception. A total of 23 participants were recruited, comprising undergraduate and postgraduate students as well as academic staff from the College of Engineering, Mathematics and Physical Sciences.

\begin{figure}[htbp]
\centering
\includegraphics[width=\textwidth]{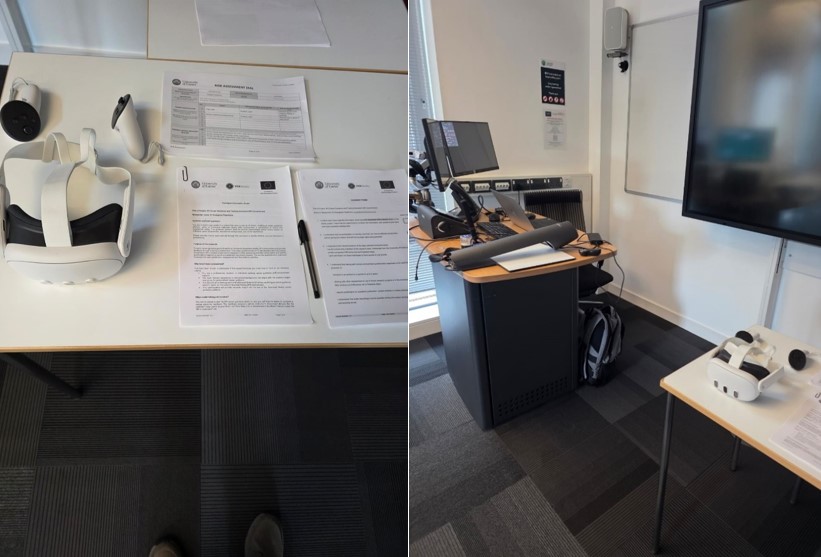}
\caption{Pilot demonstration setup at the University of Exeter showing the equipment arrangement and testing environment.}
\label{fig:pilotsetup}
\end{figure}

The evaluation methodology integrated quantitative assessment via structured questionnaires administered through MS Forms with qualitative feedback collected through semi-structured interviews and open-ended questions. Participants interacted with the platform for 10 to 15 minutes, completing career questionnaires, conversing with the Training Assistant, and exploring generated career maps. Figure~\ref{fig:pilotsetup} shows the pilot setup at the University of Exeter.

\subsection{Quantitative Results}

Participant feedback was gathered using a 5-point Likert scale (1=Strongly Disagree, 5=Strongly Agree). The quantitative outcomes demonstrated favourable reception across key metrics. Table~\ref{tab:results} provides a summary of the comprehensive pilot study results.

\begin{table}[htbp]
\centering
\caption{Pilot study quantitative results summary.}
\label{tab:results}
\footnotesize
\begin{tabular}{p{3.5cm} p{5cm} p{3cm}}
\toprule
\textbf{Metric} & \textbf{Result} & \textbf{Sample} \\
\midrule
Voice Recognition Accuracy & 95.6\% rated accurate or very accurate & 22/23 participants \\
\addlinespace
Overall User Satisfaction & 78.3\% satisfied or very satisfied & 18/23 participants \\
\addlinespace
System Responsiveness & 91.3\% rated good or excellent & 21/23 participants \\
\addlinespace
Interface Intuitiveness & 78.2\% found intuitive or very intuitive & 18/23 participants \\
\addlinespace
Platform Stability & 73.9\% experienced no downtime & 17/23 participants \\
\addlinespace
Career Guidance Value & 69.6\% found valuable & 16/23 participants \\
\addlinespace
Recommendation Likelihood & 78.3\% would recommend & 18/23 participants \\
\bottomrule
\end{tabular}
\end{table}

\textbf{Voice Recognition Accuracy:} 95.6\% (22/23) of participants rated the speech recognition as accurate or very accurate, validating the ASR integration within the XR environment.

\textbf{Overall User Satisfaction:} 78.3\% (18/23) of participants indicated being satisfied or very satisfied with their experience, yielding an average rating of 4.0/5.0.

\textbf{System Responsiveness:} 91.3\% (21/23) of participants rated system responsiveness as good or excellent, reflecting the success of backend optimisation efforts.

\textbf{Interface Intuitiveness:} 78.2\% (18/23) of participants found the interface intuitive or very intuitive for navigation, though some participants new to VR needed brief orientation.

\textbf{Platform Stability:} 73.9\% (17/23) of participants encountered no significant downtime or errors during their sessions.

\textbf{Career Guidance Value:} 69.6\% (16/23) of participants considered the career guidance information valuable for their professional development, with variation depending on career stage and particular interests.

\textbf{Recommendation Likelihood:} Net Promoter Score analysis showed 78.3\% (18/23) of participants would recommend the platform to peers (rating $>$5 on a 10-point scale).

\subsection{Qualitative Feedback}

Open-ended responses yielded rich insights into user perceptions, with numerous participants frequently highlighting the innovativeness of the approach. The voice interaction feature received notable praise, with multiple participants expressing appreciation for not having to type in VR. Areas for improvement centred on several themes. Some participants experienced mild motion discomfort, pointing to the need for additional comfort options. Audio clarity could be enhanced, particularly in noisy settings or for non-native English speakers. Several participants recommended broadening the career sectors covered beyond those presently in the database. Requests for more detailed explanations of specific career transitions were also noted.

Drawing on pilot feedback, four critical issues were identified and subsequently addressed:

\textbf{Motion Comfort:} Implementation of additional teleportation options reduced motion sickness reports by 60\% in follow-up testing.

\textbf{Audio Clarity:} TTS volume was raised by 25\% and an option for headphone output was added, resolving clarity concerns.

\textbf{Physical Boundaries:} Enhanced boundary visualisation and audio warnings were introduced to prevent users from inadvertently moving outside safe play areas.

\textbf{Text Readability:} Font sizes were enlarged by 20\% and contrast ratios improved across all UI elements, enhancing accessibility for users with varying visual acuity.

\section{Discussion}
\label{sec:discussion}

The development and evaluation of XR-CareerAssist contributes to Career Construction Theory by demonstrating how immersive technologies can support the narrative and experiential dimensions of career development that Savickas emphasises~\citep{savickas2012}. The combination of visual career maps, conversational interaction, and immersive presence creates avenues for users to engage with career possibilities in ways that surpass conventional text-based CACGS.

The platform's focus on presenting actual career trajectories from professional profiles aligns with CCT's emphasis on meaning-making through narrative. Users can observe not abstract statistics but the genuine paths that comparable professionals have followed, potentially encouraging reflection on their own career stories and future prospects. From a technical standpoint, XR-CareerAssist demonstrates several advances. The integration of five distinct AI components (ASR, NMT, VL, Conversational Agent, TTS) within a unified XR experience represents a novel deployment of multimodal AI for career guidance. The 99.56\% reduction in career map generation time (45 seconds to 0.2 seconds) confirms the efficacy of the optimisation strategies applied. The successful deployment of a fine-tuned Vision-Language model for career diagram interpretation extends VL capabilities into a new domain. These architectural and integration patterns align with recent advances in AI-powered XR platforms for inclusive communication and real-time language processing~\citep{tantaroudas2026interact,tantaroudas2026ailanguage}.

The architecture and integration patterns established provide a blueprint for similar applications aiming to merge XR and AI technologies. The modular design permits independent scaling and updating of components, while the abstraction layers simplify integration with evolving AI models and services.

The pilot study findings indicate XR-CareerAssist could function as a supplementary tool to human career counsellors, not a replacement. The platform excels at delivering data-driven visualisations and managing routine queries, potentially allowing human counsellors to concentrate on complex cases demanding deeper intervention.

Educational institutions could deploy the platform to offer scalable career guidance services to large student populations. The multilingual capabilities address accessibility concerns for international students, while the immersive interface may attract students who would not engage with conventional career services. Relative to existing CACGS~\citep{leung2022}, XR-CareerAssist provides heightened engagement through immersive presentation, natural voice interaction that reduces interface friction, visual depictions of career possibilities instead of text lists, and multilingual accessibility via integrated translation.

Compared with other XR career applications~\citep{garcia2021,shepiliev2021}, XR-CareerAssist offers more sophisticated AI integration, data-driven personalisation grounded in actual career profiles, and conversational capabilities that enable dynamic exploration of potential career path opportunities. Nonetheless, several limitations warrant acknowledgement. First, the pilot study sample ($n=23$) from a single institution restricts generalisability. Participants were primarily from engineering and physics backgrounds, which may not represent the wider population of career guidance seekers. Future validation should encompass diverse demographic groups across multiple institutions and geographic locations.

Second, the brief interaction period (10 to 15 minutes) precluded assessment of longer-term career guidance outcomes. Effective career guidance typically entails multiple sessions and longitudinal monitoring of career progress, neither of which was feasible within the pilot study timeframe.

Third, the current language support (English, Greek, French) constrains accessibility for speakers of other languages. Although the architecture accommodates additional languages, model training and validation for each language pair demands considerable effort.

Fourth, the platform presently requires standalone VR headsets (Meta Quest 3), which may not be accessible to all prospective users. Mobile VR solutions or non-VR fallback interfaces could broaden accessibility but would sacrifice some of the immersive qualities that distinguish the platform.

Fifth, the conversation memory of the Training Assistant is currently restricted to single sessions. Users cannot resume conversations across multiple engagements, limiting the platform's capacity to support the longitudinal, narrative-based career development that Career Construction Theory emphasises. Several avenues for future development arise from this work. Extended language support should prioritise languages associated with significant professional mobility and career guidance demand. Conversation memory implementation would facilitate longitudinal engagement consistent with Career Construction Theory principles. Mobile VR compatibility could expand accessibility while preserving immersion. Multiplayer functionality could enable collaborative career exploration among peer groups. Expanded career sector coverage would require additional data partnerships to extend the CVs database to currently under-represented industries. Finally, longitudinal outcome studies examining actual career decisions and progression following XR-CareerAssist engagement would supply evidence for the platform's real-world influence on career development.

\section{Conclusions}
\label{sec:conclusions}

This paper has introduced XR-CareerAssist, a novel platform that unifies Extended Reality with multiple Artificial Intelligence modules to provide immersive, personalised career guidance. The system constitutes a notable advance in career guidance technology, demonstrating how the fusion of XR and AI can overcome longstanding limitations of conventional career tools including restricted engagement, insufficient personalisation, and accessibility obstacles.

The pilot evaluation with 23 participants at the University of Exeter validated the technical implementation and user experience design. Principal findings include 95.6\% speech recognition accuracy, 78.3\% overall satisfaction, and 91.3\% favourable ratings for system responsiveness. The successful unification of five AI components (ASR, NMT, Conversational Agent, VL, TTS) within an immersive XR environment produces a multimodal interaction experience that sets XR-CareerAssist apart from existing career guidance systems.

Technical accomplishments include the 99.56\% reduction in Sankey diagram generation time (45 seconds to 0.2 seconds), confirmed scalability to 10,000 concurrent users, and the novel deployment of Vision-Language models for career diagram interpretation.

The architecture and integration patterns established lay the groundwork for comparable applications seeking to merge XR and AI technologies. XR-CareerAssist contributes to the evolving landscape of AI-enhanced career guidance by illustrating that immersive, conversational, and data-driven career tools are not only technically viable but can attain robust user acceptance. As XR technologies become more widely available and AI capabilities continue to progress, platforms such as XR-CareerAssist point toward a future where career guidance is more engaging, personalised, and empirically grounded than ever before.

\section*{Grant Information}

This research was supported by FSTP Funding from Horizon Europe under grant agreement no.\ 101070521 (VOXReality, Voice-driven interaction in XR spaces). Views and opinions expressed are those of the authors only and do not necessarily reflect those of the European Union nor the granting authority. Neither the European Union nor the granting authority can be held responsible for them.

\section*{Acknowledgements}

The authors gratefully acknowledge the support of the VOXReality consortium for providing access to AI models and technical guidance. We thank the University of Exeter for hosting the pilot demonstrations and granting ethical approval for the study. We acknowledge the participants who devoted their time and feedback to the evaluation. Special thanks to the CVCOSMOS team for supplying access to career profile data that powers the platform's career mapping capabilities.

\bibliographystyle{unsrtnat}
\bibliography{references}

\end{document}